\documentclass[11pt,graphicx]{article}
\usepackage{amssymb,amsmath,amsfonts}
\usepackage{graphicx}
\usepackage{graphics}
\usepackage{eepic,epsfig}
\usepackage{eepic,epsfig}
\begin{document}
\title{Vacuum polarization by topological defects in de Sitter spacetime}
\author{E. R. Bezerra de Mello \thanks{E-mail: emello@fisica.ufpb.br}\\
Departamento de F\'{\i}sica-CCEN\\
Universidade Federal da Para\'{\i}ba\\
58.059-970, J. Pessoa, PB\\
C. Postal 5.008\\
Brazil}
\maketitle      
\begin{abstract}
In this paper we investigate the vacuum polarization effects associated with a massive quantum scalar field in de Sitter spacetime in the presence of gravitational topological defects. Specifically we calculate the vacuum expectation value of the field square, $\langle\Phi^2\rangle$. Because this investigation has been developed in a pure de Sitter space, here we are mainly interested on the effects induced by the presence of the defects. 
\end{abstract}
\maketitle

\section{Introduction}
De Sitter (dS) space is one of the curved spacetime which has attracted more attention in quantum field theory during the past two decades. One of the main reason resides in the fact that it is maximally symmetric and several physical problems can be exactly solvable on this background; moreover in great number of inflationary models, approximated dS spacetime is employed to solve relevant problems in standard cosmology. During an inflationary epoch, quantum fluctuations in the inflaton field introduce inhomogeneities and may affect the transition toward the true vacuum. These fluctuations play important role in the generation of cosmic structures from inflation.

It is well known that different types of topological defects may have been formed in the early universe after Planck time by the vacuum phase
transition \cite{Kibble,V-S}. Depending on the topology of the vacuum manifold, $\cal{M}$, these are domain wall, strings, monopole and texture.
Cosmic strings and monopoles seem to be the most probably to be observed. Cosmic strings are linear topologically stable gravitational. The gravitational field produced by a cosmic string may be approximated by a planar angle deficit in the two-dimensional sub-space perpendicular to the string. Although recent observations data on the cosmic microwave background have ruled out cosmic strings as the primary source for primordial density perturbation, they are still candidate for the generation of a number of interesting physical effects such as gamma ray burst \cite{Berezinski}, gravitational waves \cite{Damour} and high energy cosmic rays \cite{Bhattacharjee}. Recently, cosmic strings have attracted renewed interest partly because a variant of their formation mechanism is proposed in the framework of brane inflation \cite{Sarangi}-\cite{Dvali}. Global monopoles are spherically symmetric topological stable gravitational defects which appear in the framework of grand unified theories. These objects could be produced as a consequence of spontaneous breakdown of a global $O(3)$ gauge symmetry to $U(1)$ \cite{BV}. The gravitational field produced by global monopoles may be approximated by a
solid angle deficit in the three-dimensional space.

The analysis of quantum vacuum fluctuations associated with matter fields in curved spacetime are relevant in a semiclassical theory of gravity \cite{BD}. The validity conditions under which the vacuum expectation values of the energy-momentum tensor, $\langle T_{\mu\nu}\rangle$, can act as sources for a semiclassical gravitational field are discussed in \cite{Ford,Singh}. In this context, as the first steps in this direction, here we shall investigate the vacuum expectation value (VEV) of the square of quantum fields, $\langle\Phi^2\rangle$, on the background of four-dimensional de Sitter spacetime considering the presence of a cosmic string and global monopole in it. 

This paper is organized as follow. In section \ref{sec2} we present backgrounds associated with the geometries under considerations and the solutions
of the Klein-Gordon equations admitting an arbitrary curvature coupling. Also we present the corresponding Wightman functions. In section \ref{sec3} we calculate the vacuum expectation values of the field squared. Finally we leave for the section \ref{sec4} our more relevant conclusions about this paper.

\section{Topological defects in de Sitter}
\label{sec2}
Although gravitational topological defects have been first introduced in a Minkowski bulk, they can also considered in dS background: 
\begin{itemize} 
\item For a cosmic string in dS spacetime, the geometry can be given by the following line element \cite{Mello1}:
\begin{equation}
ds^{2}=dt^{2}-e^{2t/\alpha }\left(dr^{2}+r^{2}d\phi^{2}+dz^{2}\right) \ ,  \label{ds1}
\end{equation}
where $r\geq 0$ and $\phi \in \lbrack 0,\ 2\pi /q]$ define the coordinates on the conical geometry. The parameter $q\geq	1$, is related with the linear mass density of the string, and the parameter $\alpha $ is related with the cosmological constant and Ricci scalar by the formulae
\begin{equation}
\Lambda =\frac{3}{2\alpha ^{2}}\ {\rm and} \ R=\frac{12}{\alpha ^{2}} \ .
\label{Lambda}
\end{equation}
\item For a global monopole in dS spacetime, the geometry can be given by the following line element \cite{Mello2}:
\begin{eqnarray}
\label{dS0}
ds^2=dt^2-e^{2t/\alpha}\left(dr^2+\beta^2r^2d\Omega^2\right) \ ,
\end{eqnarray}
where the coordinates, $x^\mu=(t, \ r, \ \theta, \phi)$, are defined in the intervals $t\in(-\infty, \infty)$, $\theta\in [0, \pi]$,  $\phi\in[0, \ 2\pi]$ and $r\geq 0$. The parameter $\beta\leq 1$, which codifies the presence of the global monopole, is associated with the energy scale where the defect is formed. For this geometry the curvature scalar reads
\begin{eqnarray}
R=\frac{12}{\alpha^2}+\frac{2e^{-2t/\alpha}(1-\beta^2)}{\beta^2r^2} \ .
\end{eqnarray} 
\end{itemize}
In addition to the synchronous time coordinates $t$, we may use the conformal time $\tau $\ defined according to
\begin{equation}
\tau =-\alpha e^{-t/\alpha }\ ,\ -\infty <\ \tau \ <\ 0\ ,  \label{tau}
\end{equation}
to express the above line elements. In this new coordinate system the line elements (\ref{ds1}) and (\ref{dS0}) read:
\begin{itemize}
\item For the cosmic string,
\begin{equation}
ds^{2}=(\alpha /\tau )^{2}\left(d\tau ^{2}-dr^{2}-r^{2}d\phi^{2}-dz^{2}\right)\ .  \label{ds-cs}
\end{equation}
\item For global monopole,
\begin{eqnarray}
ds^2=(\alpha/\tau)^2\left(d\tau^2-dr^2-\beta^2r^2 d\Omega^2\right) \ . \label{ds-gm}
\end{eqnarray}
\end{itemize}

As we can see, both line elements above are conformally related with the corresponding cosmic string and global monopole geometries by the function factor $(\alpha /\tau )^{2}$.

\subsection{Wightman function}
An important quantity to calculate vacuum polarization effects is the positive frequency Wightman function. Because in this paper we are more interested to analyse the vacuum polarization effects due to a cosmic string and a global monopole in dS space, the main objective of this subsection is to calculate the positive frequency Wightman functions associated with a quantum massive scalar field in a four-dimensional dS spacetime taking into account the presence of the defects. In order to do that we first obtain the complete set of eigenfunctions for the Klein-Gordon equation below, admitting an arbitrary curvature coupling:
\begin{equation}
(\nabla _\mu\nabla ^\mu+m^{2}+\xi R)\Phi (x)=0  \ ,  \label{KG}
\end{equation}
where $\xi $ is the arbitrary curvature coupling constant. 
\subsubsection{Cosmic string case}
The complete set of solutions of (\ref{KG}) in the coordinate system defined by (\ref{ds-cs}) is: 
\begin{equation}
\Phi _{\sigma}(x)=C_{\sigma }\eta^{3/2}H_{\nu}^{(1)}(\lambda\eta)J_{q|n|}(pr)e^{ikz+in\varphi} \ , \ \eta =\alpha e^{-t/\alpha } \ ,  \label{sol1}
\end{equation}
where $\lambda =\sqrt{p^{2}+k^2}$, $H^{(1)}_\mu$ and $J_\nu$ represent the Hankel and Bessel functions, respectively, and 
\begin{equation}
\nu =\sqrt{9/4-12\xi-m^{2}\alpha ^{2}},\ \ n=0,\pm 1,\pm 2,\ ... \ .
\end{equation}
This solution is characterized by the set of quantum numbers $\sigma \equiv (k,\ p,\ n)$, where $k\in(-\infty, \ \infty)$ and $p\in \lbrack 0,\ \infty )$. The coefficient $C_{\sigma }$ can be found by the orthonormalization condition
\begin{equation}
-i\int d^{D-1}x\sqrt{|g|}g^{00}[\Phi _{\sigma }(x)\partial _{t}\Phi _{\sigma^{\prime }}^{\ast }(x)-\Phi _{\sigma ^{\prime }}^{\ast }(x)\partial _{t}\Phi
_{\sigma }^{\ast }(x)]=\delta _{\sigma ,\sigma ^{\prime }}\ ,
\label{normcond}
\end{equation}
where the integral is evaluated over the spatial hypersurface $\tau =\mathrm{const}$, and $\delta _{\sigma ,\sigma ^{\prime }}$ represents the
Kronecker-delta for discrete index and Dirac-delta function for continuous ones. This leads to the result
\begin{equation}
	C_\sigma^2=\frac{q \ p}{16\alpha^2\pi}  \label{coef}
\end{equation}
for the normalization coefficient.

We shall employ the mode-sum formula to calculate the positive frequency Wightman function:
\begin{equation}
G(x,x^{\prime })=\sum_{\sigma }\Phi _{\sigma }(x)\Phi _{\sigma }^{\ast}(x^{\prime })\ .  \label{modesum}
\end{equation}
Substituting (\ref{sol1}), with respective coefficient (\ref{coef}), into (\ref{modesum}) we obtain
\begin{eqnarray}
G(x,x')&=&\frac{q(\eta\eta^{\prime})^{3/2}}{16\alpha^2\pi}\int_{0}^{\infty}dp\ p\ \int_{-\infty}^\infty d k\,\,e^{ik\Delta{z}} \times \nonumber \label{Green1}
\\
&&\sum_{n}e^{inq\Delta \phi }J_{q|n|}(pr)J_{q|n|}(pr^{\prime })H_{\nu}^{(1)}(\lambda \eta )[H_{\nu }^{(1)}(\lambda \eta ^{\prime })]^{\ast }\ ,
\end{eqnarray}
with $\Delta{z}={z}-{z}^{\prime }$, $\Delta \phi=\phi-\phi ^{\prime }$. 

The analysis of vacuum polarizations in dS spacetime, have been developed by many authors in literature. Here we are mainly interested
in quantum effects induced by the presence of the cosmic string. In order to investigate these effects we introduce below the subtracted Wightman
function
\begin{equation}
G_s(x,x')=G(x,x')-G(x,x')|_{q=1} \ .
\label{gsub}
\end{equation}
As the presence of the string does not change the curvature for the background manifold for $r\neq 0$, this function does not present additional divergences in the calculations of vacuum polarizations. Hence, for these points the function $G_s(x,x')$ is finite in the coincidence limit. 

In this paper we shall consider only the particular case where the parameter $q$ is an integer number.\footnote{Although being a very special situation, the results found can be used to provide the behavior on the vacuum polarization effects induced by a cosmic string for a general case.} For this case the Wightman function given in (\ref{Green1}) can be expressed by the sum of $q$ images of dS Wightman function \cite{Mello1}
\begin{eqnarray}
\label{Greenk}
	G(x',x)=\sum_{k=0}^{q-1}G_k(x',x) \ ,
\end{eqnarray}
with
\begin{eqnarray}
\label{gk}
	G_k(x',x)=\frac{\Gamma\left(\frac32+\nu\right)\Gamma\left(\frac32-\nu\right)}{8\alpha^2\pi^2(1-u_k^2)^{1/2}} P_{\nu-1/2}^{-1}(u_k) \ ,
\end{eqnarray}
being
\begin{equation}
u_{k}=\frac{\Delta x_k^2}{2\eta\eta'}=-1+\frac{\Delta{z}+r^{2}+r'^2-2rr'\cos (\Delta\phi +2\pi k/q)-(\Delta \eta )^{2}}{2\eta \eta ^{\prime }} \ ,
\end{equation}
where $P_\mu^\nu$ represents the associated Legendre function of first kind.

We can see that only the $k=0$ component of (\ref{Greenk}) is divergent in the coincidence limit, $x'\to x$. Finally we have that $G_s(x,x')$ is given by
\begin{equation}
G_s(x,x')=\sum_{k=1}^{q-1}G_{k}(x,x') \ ,
\label{Gsubk}
\end{equation}
which is finite at the coincidence limit. 

\subsubsection{Global monopole case} The complete set of solutions of (\ref{KG}) in the coordinate system defined by (\ref{ds-gm}) is: 
\begin{equation}
\Phi_{\sigma}(x)=C_{\sigma}\frac{\eta^{3/2}}{\sqrt{r}}H_{\mu }^{(1)}(\beta\omega\eta)J_{\nu_l}(\beta\omega r)Y_l^m(\theta, \ \phi) \ , \ \eta =\alpha e^{-t/\alpha } \ , \label{sol11}
\end{equation}
where 
\begin{eqnarray}
\label{order}
\mu=\sqrt{9/4-12\xi-m^{2}\alpha ^{2}} \ ,  \ \nu_l=\frac1\beta\sqrt{(l+1/2)^2+2(1-\beta^2)(\xi-1/8)} \ .
\end{eqnarray}
$Y_l^m$ represents the spherical harmonics of degree $l$. In (\ref{sol11}), $\sigma \equiv (\omega, \ l, \ m)$ denotes the set of quantum numbers, being $\omega\in \lbrack 0,\ \infty )$. The coefficient $C_{\sigma }$ can be found by the orthonormalization condition (\ref{normcond}). It reads
\begin{equation}
C_\sigma^{2}=\frac{\pi\omega}{4\alpha^2} \ .  \label{coef1}
\end{equation}
Employing the mode-sum formula (\ref{modesum}) to calculate the positive frequency Wightman function, after many intermediate steps, we obtain \cite{Mello2}:
\begin{eqnarray}
\label{Wight1}
G(x,x^{\prime })=\frac1{4\pi^2(\alpha\beta)^2}\left(\frac{\eta\eta'}{rr'}\right)^{3/2}\int_0^\infty \ dy\frac{\cosh(2\mu y)}{\sinh(v)} \sum_l(2l+1)P_l(\cos\gamma)e^{-\nu_l v} \ ,
\end{eqnarray}
where the variable $v$ is defined by
\begin{equation}
	\cosh(v)=\frac1{2rr'}(r^2+r'^2-\Delta\eta^2+4\eta\eta'\sinh^2(y)) \ .
\end{equation}

\section{The computation of $\langle \Phi ^{2}\rangle $} 
\label{sec3}
Having obtained the positive frequency Wightman functions the next steps is the calculation of the vacuum expectation value of the field square. Formally this value is obtained by computing the Wightman function at the coincidence limit:
\begin{equation}
\langle\Phi^2(x)\rangle=\lim_{x'\to x}G(x,x') \ .
\end{equation}
However this procedure provides a divergent result. To obtain a finite and well defined result we shall apply in this calculation the point-splitting renormalization method. This procedure is based upon a divergence subtraction scheme in the coincidence limit of the Wightman function. This will be done in a manifest form, by subtracting from the Wightman function the singular part of the Hadamard expansion, $G_{H}(x,x')$, before applying the coincidence limit, as shown below:
\begin{equation}
\label{Phi1}
\langle\Phi^2(x)\rangle_{Ren.}=\lim_{x'\to x}\left[G(x,x')-G_{H}(x,x')\right] \ .
\end{equation}

This calculation will be developed separately in this section specifying the physical situation under consideration.

\subsection{Cosmic string case}
According to our previous analysis we may write the complete Wightman function as the sum of a pure dS part plus the subtracted one as shown below:
\begin{equation}
G(x,x')=G_{dS}(x,x')+G_s(x,x') \ .
\end{equation}
Consequently
\begin{equation}
\label{Phi2}
\langle\Phi^{2}(x)\rangle= \langle \Phi^{2}(x)\rangle_{dS}+\langle \Phi^{2}(x)\rangle_s \ .
\end{equation}
The divergence is contained only in the part corresponding to the pure dS spacetime. The part induced by the cosmic string is finite for points outside the string.

Due to the maximal symmetry of the dS spacetime, $\langle \Phi^{2}(x)\rangle_{dS}$ does not depend on the spacetime point. In the special case of $q$ being an integer number $\langle \Phi^{2}(x)\rangle_s$ is essentially simplified and can be given by (\ref{Gsubk}):
\begin{eqnarray}
\label{Phi}
	\langle\Phi^2(x)\rangle_{s}=\frac{\Gamma\left(\frac{3}2+\nu\right)\Gamma\left(\frac{3}2-\nu\right)}{8\alpha^2\pi^2}	\sum_{k=1}^{q-1}\frac{P_{\nu-1/2}^{-1}(u^0_k)} {[1-(u_k^0)^2]^{1/2}} 
\end{eqnarray}
with
\begin{eqnarray}
\label{uk}
	u_k^0=-1+2(r/\eta)^2\sin^2(\pi k/q) \ .
\end{eqnarray}
Note that the VEV (\ref{Phi}) is a function of the ratio $r/\eta $ which is the proper distance from the string, $\alpha r/\eta $, measured in the units
of the dS curvature radius $\alpha $. 

In figure $1$ we have plotted the string induced part in the VEV of the field squared versus the ratio $r/\eta$ for minimally coupled scalar field ($\xi =0$) with $m\alpha =1$ (left panel) and $m\alpha =2$ (right panel). The numbers near the curves correspond to the values of the parameter $q$. For the left panel the parameter $\nu$ is real and for the right one this parameter is imaginary. In the latter case the oscillatory behavior of the VEV is seen at large distances from the string.

\begin{figure}[tbph]
\begin{center}
\begin{tabular}{cc}
\epsfig{figure=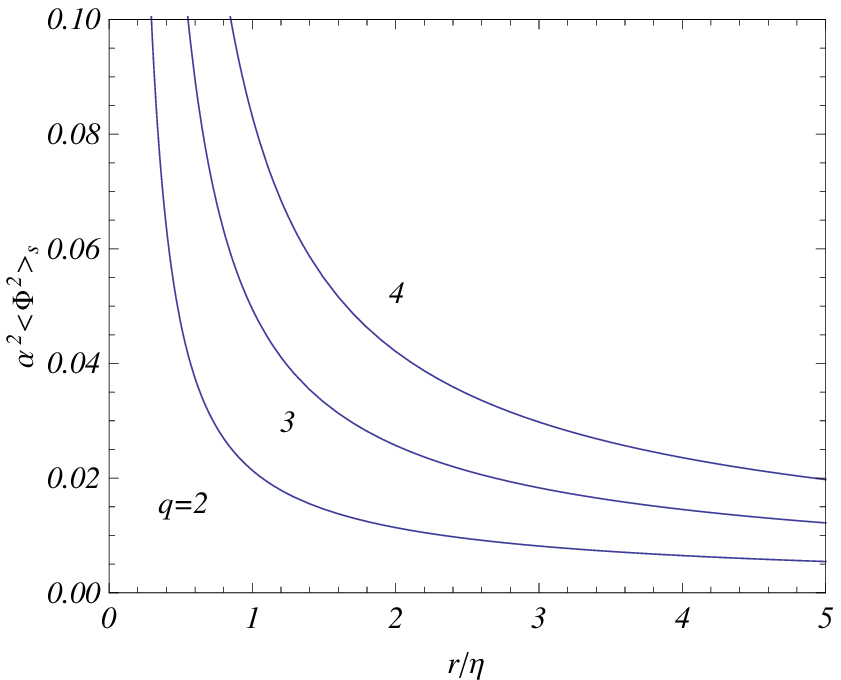,width=7.cm,height=6.cm} & \quad %
\epsfig{figure=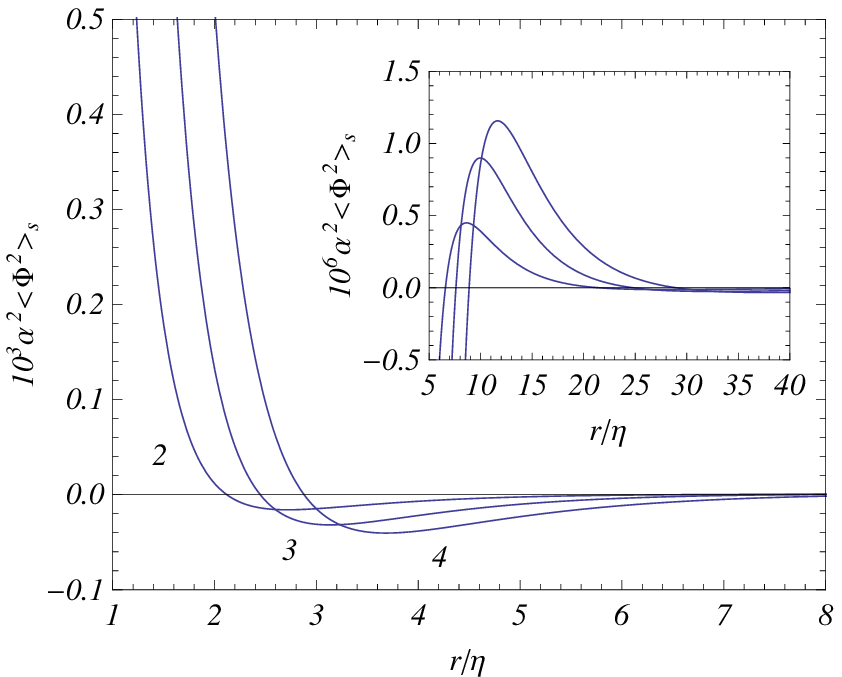,width=7.cm,height=6cm}%
\end{tabular}
\end{center}
\caption{The string induced part in the VEV of the field squared as a function of the ratio $r/\protect\eta $ for various values of the parameter $%
q$ for a minimally coupled scalar field with $m\protect\alpha =1$ (left panel) and $m\protect\alpha =2$ (right panel) in dS spacetime.}
\label{fig1}
\end{figure}

For a conformally coupled massless scalar field, $\nu=1/2$, so (\ref{Phi}) becomes
\begin{equation}
\langle \Phi ^{2}\rangle_s=\frac1{16\pi^2}\left( \frac{\eta }{\alpha r}\right)^2\sum_{k=1}^{q-1}\frac{1}{\sin^2(\pi k/q)}\ .
\end{equation}
The summation on the right-hand side of the above result can be obtained in a closed form \cite{Mello1}:
\begin{equation}
I(x)=\sum_{k=1}^{q-1}\sin ^{-2}(x+k\pi /q)=\frac{q^{2}}{\sin ^{2}(qx)}-\frac{1}{\sin ^{2}(x)}\ .  \label{IN}
\end{equation}%
Taking the limit $x\to 0$ we have $I(0)=(q^{2}-1)/3$; consequently
\begin{equation}
\langle \Phi ^{2}\rangle _{\mathrm{s}}=\frac{q^{2}-1}{48\pi ^{2}}\left(
\frac{\eta }{\alpha r}\right) ^{2}\ .  \label{D.1}
\end{equation}
The above result is an analytical function of $q$ and by analytical continuation is valid for arbitrary value of this parameter.

\subsection{Global monopole case}
In general it is not possible to provide a closed expression for the Wightman function given in (\ref{Wight1}). This fact is consequence of the non trivial dependence of the order of the Bessel function, $\nu_l$, given in (\ref{order}), with the angular quantum number $l$. With the objective to provide a more workable Wightman function which allows us to develop the calculation of the VEV of the field squared, in \cite{Mello2} is presented an approximated expression for (\ref{Wight1}) considering that the parameter $\beta$ is close to the unity.\footnote{In fact for a typical grand unified theories the parameter $\beta^2\approx 1-10^{-5}$.} In this way, taking $\gamma=0$, and developing an expansion in the parameter $\Delta^2=1-\beta^2<<1$ in $\nu_l$, up to the first order, the summation on the angular quantum number $l$ can be developed, providing for the Wightman function the expression below:
\begin{eqnarray}
\label{W4}
G(x,x^{\prime })=\frac1{16\pi^2\alpha^2\beta^2}\left(\frac{\eta\eta'}{rr'}\right)^{3/2}\int_0^\infty \ dy\frac{\cosh(2\mu y)}{\sinh^3(v/2)} \left[1-\frac{v\Delta^2}{\sinh(v)}\left(1+4\xi\sinh^2(v/2)\right)\right] \ .
\end{eqnarray}
Moreover, expanding the factor $1/\beta^2$ up to the first power in $\Delta^2$, (\ref{W4}) can be expressed as the sum of two terms, the first one corresponds to the Wightman function in a pure dS spacetime, and the second term, corresponds to the contribution coming from the presence of the monopole, as shown below:
\begin{equation}
\label{dS-gm}
G(x,x')=G_{dS}(x,x')+G_{gm}(x,x') \ ,
\end{equation}
with
\begin{eqnarray}
\label{gm1}
G_{gm}(x,x')=\frac1{16\pi^2\alpha^2}\frac{\Delta^2}{(\rho\rho')^{3/2}}\int_0^\infty \ dy\frac{\cosh(2\mu y)}{\sinh^3(v/2)}
\left[1-\frac{v}{\sinh(v)}\left(1+4\xi\sinh^2(v/2)\right)\right] \ ,
\end{eqnarray}
being
\begin{equation}
\cosh(v)=\frac{\rho^2+\rho'^2}{2\rho\rho'}+\frac2{\rho\rho'}\sinh^2(y) \ .
\end{equation}
In the above expressions we have introduced the dimensionless coordinate $\rho=r/\eta$, which corresponds to the proper distance from the monopole, $\alpha r/\eta$, measured in units of dS curvature radius, $\alpha$.

As consequence of this procedure, the VEV of the field squared, (\ref{Phi1}), is given as the sum of two distinct contributions. The first one corresponds the value in a pure dS spacetime, and the second contribution is consequence of the presence of the global monopole:
\begin{equation}
\label{Pgm}
\langle\Phi^{2}(x)\rangle_{Ren} =\langle\Phi^{2}(x)\rangle_{dS}+\langle\Phi^{2}(x)\rangle_{gm} \ .  
\end{equation}

Because the global monopole modifies the curvature of the background geometry introducing new divergences at the radial coordinate coincidence limit, $r'=r$, in the Wightman function, a finite and well defined contribution to the last term of (\ref{Pgm}) is given by: 
\begin{equation}
\label{Pgm1}
\langle \Phi^{2}(x)\rangle_{gm}=\lim_{x'\to x}\left[G_{gm}(x,x')-{\tilde{G}}_{H}(x,x')\right] \ , 
\end{equation}
where ${\tilde{G}}_{H}(x,x')$ corresponds to the expansion in the Hadamard function due only to the presence of the monopole.

In \cite{Chris1} is provided a general expression for the singular expansion of the Hadamard function. It is given in terms of the one-half of the geodesic distance between two points $x$ and $x'$. It reads:
\begin{eqnarray}
G_H(x,x')=\frac1{8\pi^2\sigma(x,x')}+\frac1{16\pi^2}\left[m^2+\left(\xi-\frac16\right)R\right]\left[2\gamma+\ln\left(\frac{m^2_{DS}\sigma(x,x')}2\right)\right] \ ,
\end{eqnarray}
where $m_{DS}$ is equal to $m$ for massive scalar field, and equal to $\mu$, an arbitrary infrared cutoff energy scale, for a massless field. The curvature scalar $R$ corresponds to dS one, $R_{dS}=\frac{12}{\alpha^2}$, plus the correction due to the presence of the monopole, $R_{gm}=2\frac{(1-\beta^2)\eta^2}{\alpha^2\beta^2r^2}$. In our approximation we may use $R_{gm}\approx 2\frac{\Delta^2\eta^2}{\alpha^2r^2}$. Consequently the term in the Hadamard function needed to renormalize (\ref{Pgm1}) is:
\begin{equation}
{\tilde{G}}_{H}(x,x')=\frac{\Delta^2}{8\pi^2\alpha^2}\frac{1}{\rho^2}\left(\xi-\frac16\right)\left[2\gamma+\ln\left(\frac{m^2\sigma(x,x')}2\right)\right] \ .
\end{equation}

Substituting into the above expression, the radial one-half of the geodesic distance given by $\sigma(x,x')=\frac{\alpha^2\Delta r^2}{2\eta^2}$, we obtain
\begin{equation}
\label{H1}
{\tilde{G}}_{H}(x,x')=\frac{\Delta^2}{4\pi^2\alpha^2}\frac{1}{\rho^2}\left(\xi-\frac16\right)\ln\left(\frac{m\alpha\Delta\rho}2\right)+\frac{\Delta^2}{4\pi^2\alpha^2}\frac{\gamma}{\rho^2}\left(\xi-\frac16\right) \ .
\end{equation}
At this point it is convenient to express the logarithmic singular contribution in (\ref{H1}) by an integral form. This can be done by using the Legendre function of second kind $Q_0$, in an integral representation \cite{Grad}:
\begin{equation}
\label{Int.Repr}
Q_0(\cosh (u))=\frac1{\sqrt{2}}\int_u^\infty \frac{dy \ e^{-y/2}}{\sqrt{\cosh(y)-\cosh(u)}} \ ,
\end{equation}
being $\cosh(u)=\frac{\rho^2+\rho'^2}{2\rho\rho'}$. Doing this we find
\begin{eqnarray}
\label{H2}
{\tilde{G}}_{H}(x,x')&=&\frac{\Delta^2}{4\pi^2\alpha^2}\frac{1}{\rho^2}\left(\xi-\frac16\right)\ln\left(\frac{m\alpha(\rho+\rho')}2\right)+\frac{\Delta^2}{4\pi^2\alpha^2}\frac{\gamma}{\rho^2}\left(\xi-\frac16\right)\nonumber\\
&-&\frac{\Delta^2}{4\pi^2\alpha^2}\frac{1}{\rho^2}\left(\xi-\frac16\right)Q_0(\cosh(u)) \ .
\end{eqnarray}

Now substituting (\ref{gm1}) and (\ref{H2}) into (\ref{Pgm1}), and after some intermediate steps, we obtain:
\begin{eqnarray}
\label{PhiD4}
\langle \Phi^{2}(x)\rangle_{gm}&=&\frac{\Delta^2}{16\pi^2\alpha^2}\int_0^\infty  dy  \left\{\frac{\cosh(2\mu y)}{\sinh^3(y)}\left[1- \frac{{\tilde{v}}}{2\sinh(y)\sqrt{\rho^2+\sinh^2(y)}}\left(\rho^2+4\xi\sinh^2(y)\right)\right]\right.\nonumber\\
&+&\left.\frac{2(\xi-1/6)}{\rho^2}\frac{e^{-y/2}}{\sinh(y/2)}\right\}-\frac{\Delta^2}{4\pi^2\alpha^2}\frac1{\rho^2}\left(\xi-\frac16\right)
\ln(m\alpha\rho)-\frac{\Delta^2}{4\pi^2\alpha^2}\frac\gamma{\rho^2}\left(\xi-\frac16\right) 
\end{eqnarray}
where
\begin{equation}
\label{v}
{\tilde{v}}=2 \ {\rm arcsinh}\left(\frac{\sinh(y)}{\rho}\right) \ .
\end{equation}
As we can see, the global monopole induced part in the VEV of the field squared, presents divergences at the monopole's position, given explicitly by the terms proportional to $\frac1{\rho^2}$ and $\frac{\ln(m\alpha\rho)}{\rho^2}$. Unfortunately it is not clear the behavior in (\ref{PhiD4}) given by the integral contribution. So, in order to exhibit this behavior, in figure $2$ we have plotted this contribution as function of the ratio $\rho=r/\eta$ (proper distance from the monopole measured in units of $\alpha $) for minimally coupled scalar field ($\xi =0$), with $m\alpha =1$, represented by points, and $m\alpha=\sqrt{2}$, by solid line, in the left panel; and with $m\alpha =\sqrt{14}/2$, by points, and $m\alpha=\sqrt{10}/2$, by solid line, in the right panel. For the left panel the correspondent values for $\mu$ are reals and for the right panel they are imaginaries. Also by this figure we can observe that for $\rho\to 0$, all the integrals become large and negative, indicating a possible divergence. In fact analysing the integrand of (\ref{PhiD4}) for small values of $\rho$ we observe a divergence $\frac1{\rho^2}$, which vanishes for $\xi=1/6$, plus a logarithmic divergent term, $\ln(\rho)$. 

\begin{figure}[tbph]
\begin{center}
\begin{tabular}{cc}
\epsfig{figure=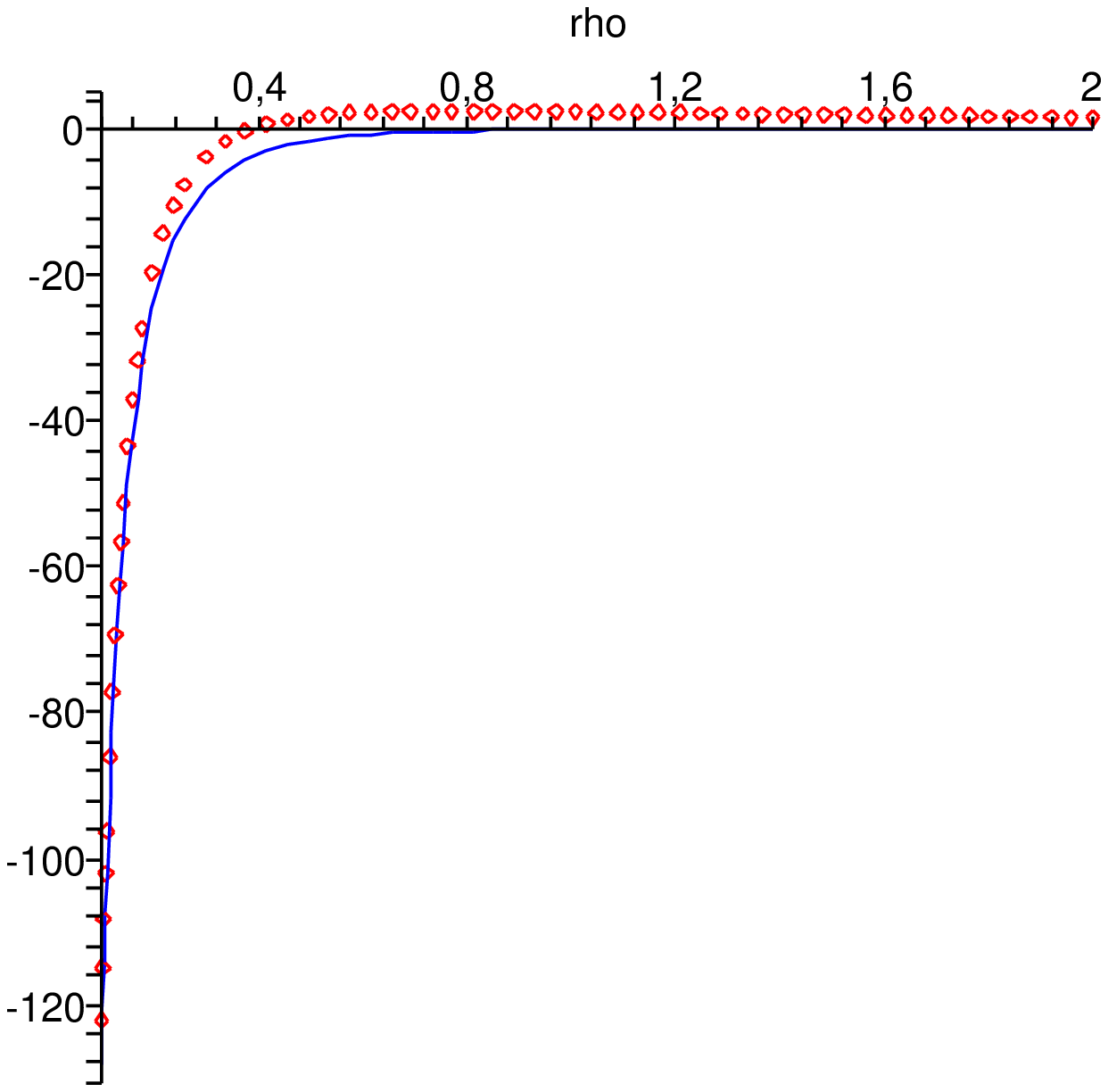,width=7.cm,height=6.cm} & \quad %
\epsfig{figure=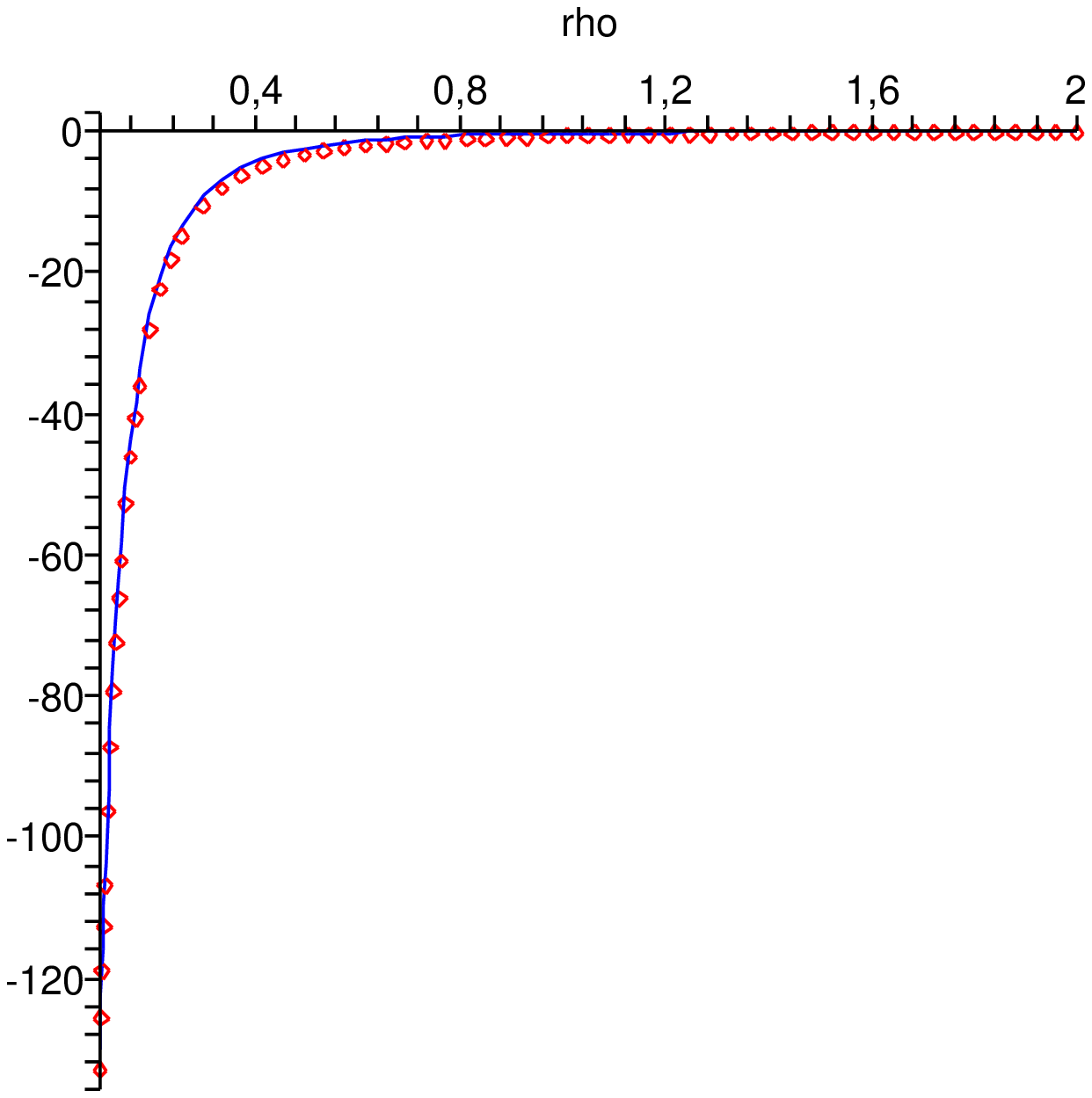,width=7.cm,height=6cm}%
\end{tabular}
\end{center}
\caption{The behavior of the integral contribution induced by the global monopole in the VEV of the field squared, Eq. (\ref{PhiD4}), as a function of the ratio $\rho=r/\eta $ for a minimally coupled scalar field with $m\alpha =1, \ \sqrt{2}$ (left panel), and $m\alpha =\sqrt{14}/2, \ \sqrt{10}/2$ (right panel). For the first values, the behavior are represented by points, and for the seconds ones by solid lines.}
\label{fig1}
\end{figure}

\section{Conclusions}
\label{sec4}
In this paper we have revisited the quantum effects associated with massive scalar fields in four-dimensional dS space in presence of two specific gravitational topological defects: $i)$ an idealized cosmic string and  $ii)$ point-like global monopole. In fact in these analysis we were more interested to calculate the contributions induced by the gravitational defects themselves on the VEV of the field square. The reason is because the vacuum polarization effects associated with a quantum scalar quantum fields in dS spacetime have been investigated by many authors in literature, and the corresponding results do not depend on the spacetime point. In this way we can compare the influence due to the defects on the vacuum polarization effects with the dS contribution.  

In order to develop these analysis we have presented the complete Wightman functions as the sum of two contributions: $i)$ The first ones correspond to the Wightman function on the bulk in the absence of the defects and $ii)$ the second ones are induced by the presence of the defects. As consequence of this formalism, we have obtained, for the VEV of the field square, the following result:
\begin{equation}
\label{PP}
\langle\phi^2\rangle=\langle\phi^2\rangle_{dS}+\langle\phi^2\rangle_d \ ,
\end{equation}
where
\begin{eqnarray}
\label{PdS}	\langle\Phi^2(x)\rangle_{dS}&=&\frac1{8\pi^2\alpha^2}\left\{\left(\frac{m^2\alpha^2}2+6\xi-1\right)\left[\Psi\left(\frac32+\nu\right)+\Psi\left(\frac32-\nu\right)-2\ln(m\alpha)\right]\right.\nonumber\\
&-&\left.\frac{(6\xi-1)^2}{m^2\alpha^2}+\frac1{30m^2\alpha^2}-6\xi+\frac23\right\} \ ,
\end{eqnarray}
corresponds to the dS spacetime contribution,\footnote{The $\Psi(x)$ in (\ref{PdS}) corresponds the logarithmic derivative of the gamma-function.} and the second term on the right hand side of (\ref{PP}), $\langle\phi^2\rangle_d$, corresponds to the induced gravitational defects contributions. By the results obtained in (\ref{Phi}) and (\ref{PhiD4}), for the latter, the corresponding results become more relevant than (\ref{PdS}) near the defects. At large distance the situation is the opposite, $\langle\phi^2\rangle_{dS}$ becomes dominant.

The results found for $\langle\phi^2\rangle_d$ are divergent at the defects' position, i.e. $r\approx 0$. These facts are consequence of the idealized models adopted for them. In a realistic point of view, cosmic strings and global monopoles have a non-vanishing radius. Taking into account a nontrivial inner structure for these defects, the VEV associated with quantum fields should be considered in both, external and internal regions, separately. Regarding to this fact, considering these defects in a Minkowski bulk, the VEV associated with scalar field in global monopole spacetime have been analysed in \cite{Mello3,Mello4} and \cite{Mello5} for scalar and fermionic fields, respectively. In these analysis we have obtained that the vacuum polarization effects in the region inside the defect are regular at $r\approx 0$. We expect that assuming a more realistic model for cosmic string and global monopole in dS spacetime, similar results will also be found. 

\section*{Acknowledgments}
E.R.B.M. thanks Conselho Nacional de Desenvolvimento Cient\'{\i}fico e Tecnol\'{o}gico (CNPq) for partial financial support, FAPESQ-PB/CNPq (PRONEX) and
FAPES-ES/CNPq (PRONEX).

\end{document}